\documentclass{PoS}

\title{Phase correction of VLBI with water vapour radiometry}

\ShortTitle{VLBI Phase Correction with WVR}

\author{\speaker{Alan. L. Roy}\thanks{We thank Clemens Simmer (Dept of
    Meteorology, Uni. Bonn) for carrying out the radiosonde launches at
    Effelsberg and for analyzing the data to derive zenith path delays.  We
    had a lot of fun.  We thank also Galina Dick and her colleagues at
    GeoForschungsZentrum, Potsdam for analyzing the GPS data to derive zenith
    path delays for comparison with those from the radiometer.  We thank also
    Ivan Agudo for providing us with the 86\,GHz VLBI data on NRAO\,150 with
    which to test the WVR phase corrections.}\\
  Max-Planck-Institut f\"ur Radioastronomie, Auf dem H\"ugel 69, 53121
  Bonn, Germany\\
  E-mail: \email{aroy@mpifr-bonn.mpg.de}}

\author{{Helge Rottmann}\\
        Max-Planck-Institut f\"ur Radioastronomie, Auf dem H\"ugel 69, 53121
        Bonn, Germany\\
        E-mail: \email{rottmann@mpifr-bonn.mpg.de}}

\author{{Ute Teuber}\\
        Max-Planck-Institut f\"ur Radioastronomie, Auf dem H\"ugel 69, 53121
        Bonn, Germany\\
        E-mail: \email{uteuber@mpifr-bonn.mpg.de}}

\author{{Reinhard Keller}\\
        Max-Planck-Institut f\"ur Radioastronomie, Auf dem H\"ugel 69, 53121
        Bonn, Germany\\
        E-mail: \email{rkeller@mpifr-bonn.mpg.de}}
      
      \abstract{We demonstrate phase correction of 3\,mm VLBI observations
        using the scanning 18\,GHz to 26\,GHz water vapour radiometer at
        Effelsberg and we demonstrate an absolute accuracy of 15\,mm in zenith
        path delay by comparing with GPS and radiosondes.  This accuracy
        should provide significant improvement in astrometric phase
        referencing observations.  It is not good enough for geodetic VLBI to
        replace the tropospheric delay estimation but could be used to remove
        short-term path-length fluctuations and so improve the geodetic
        observables.  We discuss lessons learned and opportunities for further
        improvement.}

\FullConference{The 8th European VLBI Network Symposium on New Developments in VLBI Science and Technology
                and EVN Users Meeting  \\
                September 26-29 2006\\
                Torun, Poland}

\begin{document}

\section{Introduction}

The Effelsberg radio telescope was equipped in March 2004 with a 22\,GHz 
water vapour
radiometer for phase correction of tropospheric
phase fluctuations during VLBI experiments at frequencies up to 86\,GHz.
The radiometer also measures tropospheric path delay for astrometric and
geodetic VLBI experiments.  It has been used during routine VLBI experiments
at 86\,GHz for validation tests and software is being developed to provide the
data routinely to observers.

VLBI at 86\,GHz yields resolution down to 60\,$\mu$as with global
baselines.  Experiments up to 230\,GHz have produced fringes with a record
smallest spacing of 30\,$\mu$as, on the 6.4\,G$\lambda$ trans-atlantic baseline
between Pico Veleta and the HHT (Krichbaum et al. 2004),
which is comparable to the 27\,$\mu$as diameter predicted by Falcke, Melia \&
Agol (2000) for the silhouette of the event horizon in Sgr
A$^{*}$.  However, the signal-to-noise ratios of the 230\,GHz detections were
only 7.3 on on 3C 454.3 and 6.4 on 0716+714 due to a combination of
decoherence due to atmospheric phase fluctuations during the 7 min
integration, relatively high system noise, and atmospheric opacity.
Sensitivity could be improved by a factor of two to four by 
correcting for the coherence
loss due to atmospheric phase noise using water vapour radiometry.  Water 
vapour radiometry
monitors with high precision the strength of emission from atmospheric water
vapour and infers a time-dependent correction to the
VLBI phase, which then improves the temporal coherence.

\section{The Water Vapour Radiometer at Effelsberg (WAVE)}

We have built a scanning 18\,GHz to 26\,GHz water vapour radiometer and
installed it on the focus cabin roof at Effelsberg viewing along the optical
axis.  It is based on a prototype by Alan Rogers that was demonstrated at BIMA
and during 3\,mm VLBI on the baseline between BIMA and Kitt Peak (Tahmoush \&
Rogers 2000).  The radiometer is uncooled to reduce cost, is scanning to
reduce the parts count and hence cost and to provide better stability since
offsets and drifts affect all channels and so are common-mode noise sources,
which are rejected to some extent when differencing on-line and off-line
channels.  We paid particular attention to RF shielding, weather proofing,
temperature stabilization using a Peltier element to maintain the RF
electronics near 30$^{\circ}$C and we added a noise diode for noise-adding
radiometry to monitor the system gain.  Monitoring and control takes place
over a TCP/IP connection on optical fibre, so the system requires only power
and a network connection to operate.  The data are logged into a MySQL database
from where they can be examined and downloaded through a web interface.

Photographs, schematics, measured performance and
correction of atmospheric opacity have been shown by Roy et al.
(2004).  Here, we demonstrate the correction of tropospheric phase
fluctuations during an 86\,GHz VLBI experiment and we
validate the absolute calibration for astrometric and geodetic application by
comparing to GPS and radiosonde measurements.

\section{Demonstration of Phase Correction}

A typical water line spectrum measured by the radiometer is shown in Fig 1a,
as presented by the web interface to the WVR database.  The line is centred on
22.2\,GHz and is pressure-broadened to $\sim 3$\,GHz FWHM.  The line appears
asymmetric as it sits on the frequency-squared wing of the oxygen line at
60\,GHz.  Such a spectrum is measured each 6\,s.  We fit to each spectrum a
three-component model consisting of a frequency-squared baseline underneath
the water line plus a constant offset and a van
Vleck-Weisskopf profile to the line to estimate the strength of the line above
the baseline.  From the line strength we infer the path length using the
theoretical relationship between refractive index and emissivity of water
vapour derived by Tahmoush \& Rogers (2000).

\begin{figure}
\centering
\includegraphics[width=15.2cm]{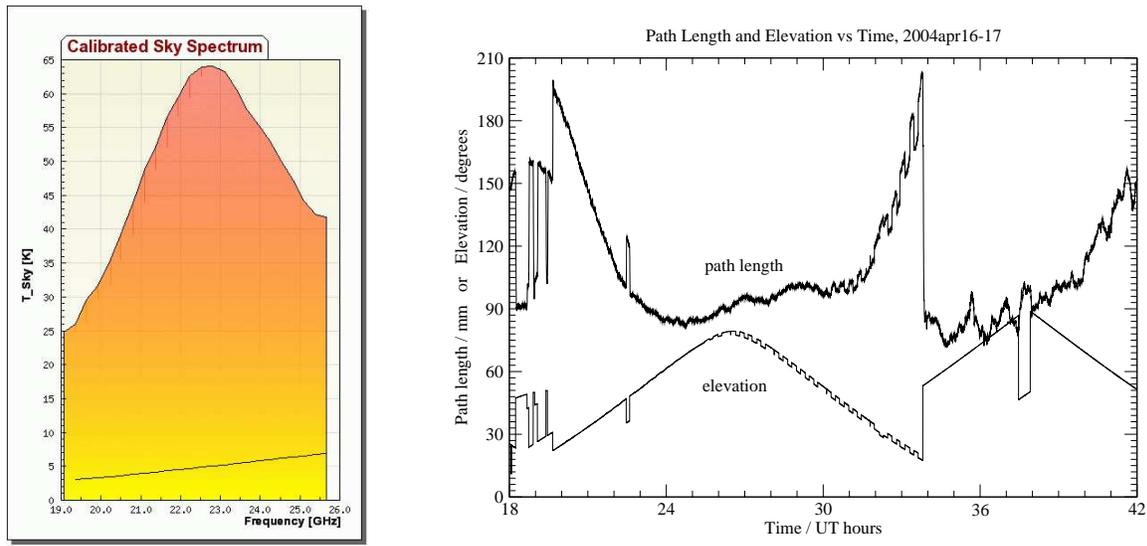}
\caption{
  {\bf a) (left):} An example 22\,GHz water line spectrum measured with the water
  vapour radiometer at Effelsberg on 2006jul20 11:56 UT at an elevation of
  54$^{\circ}$ under cloud-free sky.  The line curving upwards at the bottom
  is one of three components (the frequency-squared baseline) that was fit to
  the spectrum.
{\bf b) (right):} An example time series of tropospheric line-of-sight path length
derived from the line strength measurements made with the water vapour
radiometer at Effelsberg on 2004apr16/17 during an 86\,GHz VLBI experiment.}
\label{spectrum}
\end{figure}

A typical time series of tropospheric path length inferred from the WVR is
shown in Fig 1b.  This shows a 24\,h period during the 86\,GHz VLBI experiment
on 2004apr16 to 17.  The path length varies with elevation as the
airmass changes and the short-term path-length fluctuations grow throughout
the period due to a change in the weather.

A 7\,min piece of the time series is shown in Fig 2a along with the
measured VLBI phase on the baseline Effelsberg to Pico Veleta.  The inferred
path length and the observed VLBI phase correlate strongly.  Correcting
the VLBI phase with the WVR corrections reduced the path rms from 1.0\,mm
rms to 0.47\,mm rms and raised the coherence from 0.45 to 0.86 for a 240\,s
time-scale.  Phase correction data from Pico Veleta from the
230 GHz continuum radiometer described by Bremer (2002) were applied, further
reducing the rms path fluctuations to 0.34 mm and raising the coherence to
0.9.  In total the coherent signal-to-noise ratio improved by a factor of 2.1.

\begin{figure}
\centering
\includegraphics[width=15.2cm]{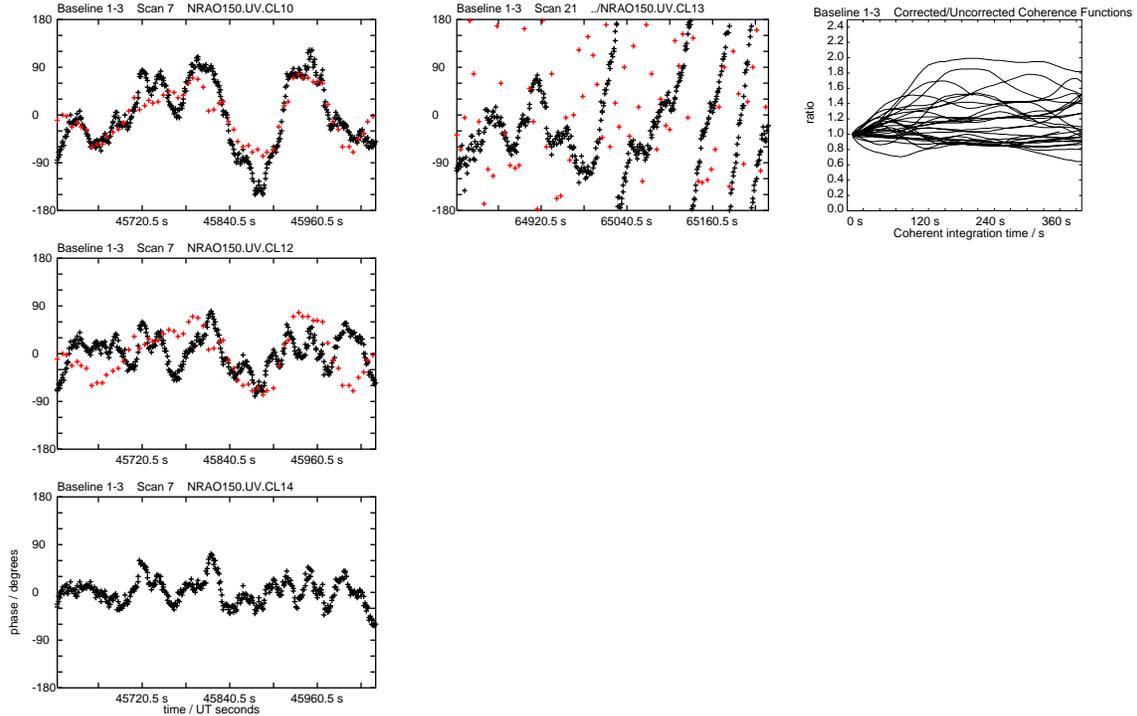}
\caption{
  {\bf a) (left column):} Phase correction of a 7\,min long 86\,GHz VLBI scan
  on NRAO\,150 on the baseline Pico Veleta - Effelsberg on 2004apr17 at
  12.7\,h UT.  Left top: the VLBI phase time series in black and the
  WVR-derived phase correction in red show good correlation.  Left middle: the
  VLBI phase time series after applying the Effelsberg WVR correction shows
  reduced phase fluctuations.  Left bottom: phase corrections at Pico Veleta
  based on the measured continuum brightness at 230\,GHz made in parallel with
  the 86\,GHz VLBI observations are applied to the middle plot, further
  reducing the phase fluctuations.  The rms path fluctuations reduced from 1.0
  mm to 0.34 mm and the coherence rose from 0.45 to 0.9 over a 240 s
  time-scale.  {\bf b) (centre column):} Phase correction of another 86
  GHz\,VLBI scan on NRAO\,150 on the baseline Pico Veleta - Effelsberg on
  2004apr17 at 18.0\,h UT this time affected by cloud emission that was not
  separated and so contaminated the path length measurement.  {\bf c) (right
    column):} The ratio of coherence after and before phase correction is
  plotted vs coherent integration time for all scans on NRAO\,150 on
  2004apr17, excluding those affected by cloud based on the high rms in the
  time series.  A ratio > 1 indicates improvement.  Most scans were improved
  by the phase correction.  }
\label{phaseCorrection1}
\end{figure}

%

Another scan during cloud is shown in Fig 2b.  The phase
corrections varied dramatically and no phase correlation could be seen.  We
are now improving the cloud separation.  
Phase correction in the presence of cloud has been reported by Bremer (2006)
at Plateau de Bure.

The fraction of scans for which the WVR corrections improved the coherence is
summarized in Fig 2c, which shows the coherence after correction divided by
the coherence before correction.  If the correction makes an improvement, the
ratio is greater than unity.  Most scans lie between 1 and 2, with the worst
case being 0.75.  Scans affected by cloud were removed for this
summary.


Corrections applied to four subsequent 86\,GHz VLBI experiments yielded no
further cases of improved coherence; in these four experiments applying WVR
corrections increased the rms phasse noise, unfortunately.  The cause is under
investigation; perhaps the atmospheric stability during those experiments was
already very good and the WVR noise raised the rms.  This is supported since
the WVR corrections were often seen to be stable and the VLBI phase wandered
by rather more suggesting that the phase noise was dominated by the other end
of the baseline.  Other observatories have also seen periods when the phase
corrections from WVRs fail to reduce the tropospheric phase noise at the VLA
(Chandler et al. 2004), Plateau de Bure (Bremer 2006), and OVRO (Woody,
Carpenter \& Scoville 2000) though not as often as in our tests.

\section{Validation of Absolute Calibration}

Astrometric and geodetic applications of water vapour radiometry require good
absolute calibration; low thermal noise is less important.  To check the
absolute calibration, we carried out an intercomparison of WVR, five GPS
stations closest to Effelsberg out to 120 km and three radiosonde launches on
2005jul27.  The GPS data were analyzed at GeoForschungsZentrum, Potsdam and
the radiosonde data were analyzed by C. Simmer at the Dept of Meteorology,
Uni. Bonn to derive zenith tropospheric path delays.  We found agreement
within 15 mm between the WVR, all GPS receivers and three radiosonde launches
(Fig 3).

\begin{figure}
\centering
\includegraphics[width=10cm]{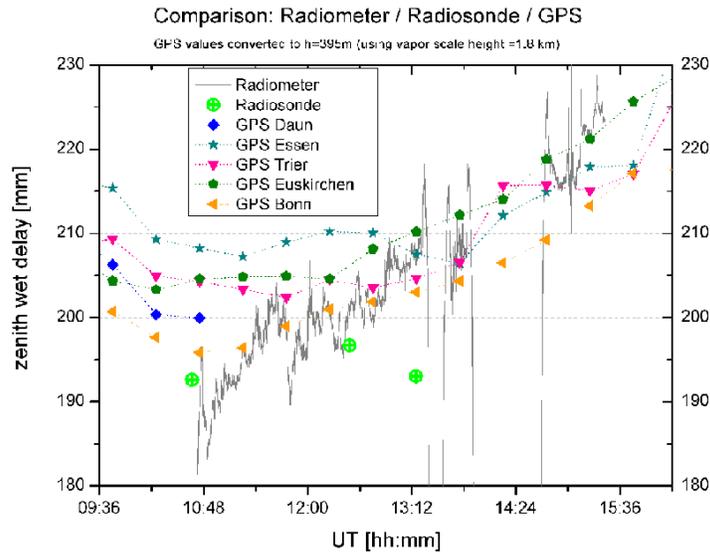}
\caption{
  Time series of zenith wet delay measured by the Effelsberg WVR, GPS
  receivers between 18\,km and 120\,km distance from the Effeslberg telescope,
  and radiosondes launched 400\,m ENE of the telescope.  The three techniques
  agree typically within 15\,mm.  The third radiosonde profile (at 13:12\,UT)
  was incomplete due to interrupted data reception.  Hence, this one
  measurement has larger uncertainty. }
\label{intercomparison}
\end{figure}

\section{Lessons Learned}

We came to appreciate the critical importance of good temperature
stabilization for maintaining absolute calibration and maintaining the
$3\times10^{-4}$ short-term gain stability required for phase correction.  We
had initially 0.7\,$^{\circ}$C variation in the internal temperature over a
24\,h period, however superimposed on this was a temperature oscillation with
a 3\,min period and peak-to-peak amplitude of 20\,mK due to the time lag for
heat flow from the Peltier element to the temperature sensor located beside
the LNA.  The temperature oscillation caused the detector output power to vary
by 250\,mK peak-to-peak which was enough to degrade the phase correction.  We
stopped the oscillation by weakening the thermal contact between the Peltier
element and the RF components by reducing the speed of a fan and removing
copper straps but the internal temperature then followed the external
temperature more closely, producing 6\,$^{\circ}$C or more temperature changes
over periods of days.  Much tighter temperature regulation has been
demonstrated by, for example, Tanner (1998).

The frequency spanned by the radiometer might be too narrow to allow
separation of cloud emission with the precision required to allow phase
correction to continue during periods of heavy cloud.  A channel on the
continuum at 50 GHz would help this a lot (Crewell 2006).  However, phase
correction in the presence of clouds with a narrower frequency span than ours
has been demonstrated by Bremer (2006) at Plateau de Bure, thus proving the
principle.

The spillover contribution is unexpectedly large (8\,K)
since the feedhorn over-illuminates the dish at low frequencies.  A new
horn with larger edge taper will alleviate this source of noise.

The noise diode proved somewhat unstable, and the system gain
variations could be better removed using the temperature measured near the
LNA along with the temperature coefficient of the
amplifier.  Methods to improve the noise diode stability have been described
by Tanner (1998).

\section{Conclusion}

The Effelsberg water vapour radiometer corrections have been demonstrated to
improve phase coherence during one high-frequency VLBI experiment, though not
in four subsequent experiments.  The tropospheric delay measurements agreed
with 15\,mm accuracy when compared to GPS and radiosondes.

\end{document}